\documentclass{article}

\usepackage[left=2cm,top=2.5cm,right=2cm,nohead,nofoot]{geometry}
\usepackage{mathrsfs,mathpazo,avant,tabularx}
\usepackage[utf8]{inputenc}

\usepackage[table]{xcolor}
\usepackage{colortbl,authblk}
\usepackage{stmaryrd}
\usepackage{amssymb,amsmath,amsthm,amsfonts,amsbsy}
\usepackage{bm,bibunits,color,chngcntr,epsfig,epstopdf,graphicx,dsfont}
\usepackage{hyperref,lipsum,,makecell,mathrsfs,rotating}
\usepackage[english]{babel}
\usepackage[normalem]{ulem}
\usepackage{indentfirst}

\newcommand{\be}{\begin{equation}}
\newcommand{\ee}{\end{equation}}
\newcommand{\bea}{\begin{eqnarray}}
\newcommand{\eea}{\end{eqnarray}}
\newcommand{\ket}{\rangle}
\newcommand{\bra}{\langle}

\newcommand{\I}{\mathds{1}}
\newcommand{\ra}{\rightarrow}

\def\C#1{\mathcal #1}

\newcommand{\T}[2]{\textsf{#1#2}}

\definecolor{gray}{gray}{0.9}

\begin{document}
\newtheorem{theorem}{Theorem}
\newtheorem{prop}[theorem]{Proposition}
\newtheorem{corollary}[theorem]{Corollary}
\newtheorem{open problem}[theorem]{Open Problem}
\newtheorem{definition}{Definition}
\newtheorem{remark}{Remark}
\newtheorem{example}{Example}
\newtheorem{task}{Task}

\title{A comparative study of universal quantum computing models: towards a physical unification}
\author{Dong-Sheng Wang \thanks{Email: \href{mailto:wds@itp.ac.cn}{wds@itp.ac.cn}; CAS Key Laboratory of Theoretical Physics, Institute of Theoretical Physics,
Chinese Academy of Sciences, Beijing 100190, China}}

\maketitle


\begin{abstract}
    Quantum computing has been a fascinating research field in quantum physics.
    Recent progresses motivate us to study in depth the universal quantum computing models (UQCM),
    which lie at the foundation of quantum computing and have tight connections with fundamental physics. 
    Although being developed decades ago,
    a physically concise principle or picture to formalize and understand UQCM is still lacking.
    This is challenging given the diversity of still-emerging models,
    but important to understand the difference between classical and quantum computing.
    In this work, we carried out a primary attempt to unify UQCM by classifying a few of them
    as two categories, hence making a table of models. 
    With such a table, some known models or schemes appear as hybridization or combination of models,
    and more importantly, it leads to new schemes that have not been explored yet.
    Our study of UQCM also leads to some insights into quantum algorithms.
    This work reveals the importance and feasibility of systematic study of computing models. 
\end{abstract}

\section{Introduction}
\label{sec:int}

The development of quantum computing has been profound.
The very concept of quantum computing was originally established by the study of \emph{modeling},
either based on Hamiltonian or circuits~\cite{Ben80,Fey82,Deu85,Mar90},
and the quantum circuit model is the most well-known one~\cite{NC00,KSV02}.
A modeling mainly refers to the framework of formalizing the way of information processing,
which is an important subject for computation~\cite{Sav98,AB09},
and it closely relates to other tasks such as information encoding and problem solving. 

In the quantum circuit model (QCM), a computation is carried out by a sequence of gates, each acting on a small number of qubits.
Although being quantum, there are classical degree of freedoms in it 
such as the spacetime location of gates. 
The program describing a gate sequence is also often classical. 
Attempts to shift the boundary between the quantum and classical parts have been made,
and it proves difficult to be fully quantum~\cite{Gru99,Mye97,NC97}.
At the meantime, various universal quantum computing models (UQCM) 
have been developed~\cite{Kit03,Wat95,KW97,SW04,RB01,FGG+01,AVK+08,VWC09,BFK09,Long06,Long11,CAPV13,CGW13},
and some of them hardly have classical analog.

Quantum computing has proven to be powerful and 
understanding the source of its power requires sustained efforts.
A recent breakthrough is the formal unification of many quantum algorithms via
quantum singular value transformation (QSVT)~\cite{GSLW19,MRTC21}.
This motivates us to systematically study UQCM and seek a unification of them.
From a physical point of view, our study confirms this with a unifying framework of describing UQCM.
In this work, we classify UQCM according to the ways of information processing and information protection.
Therefore, we identify two categories of them and this leads to the table of UQCM (Tab.~\ref{tab:hym}).
The three category-I models represent quantum states carrying information in different ways,
while the five category-II models stand for distinct ways to manipulate logical information.
This yields at least fifteen universal schemes,
some of which have not been explored yet,
and there are also vast rooms for hybrid and other ones. 

The unification of UQCM is made possible due to a few recent progresses. 
Despite the long history of quantum Turing machine (QTM)~\cite{Deu85,BV97,Yao93},
its functionality remains rather illusive due to its sophisticated configuration~\cite{Mye97,Oza98,Shi02,For03,MW19,Wang20,GMM20}.
A simplification of it using locality was proposed recently~\cite{Wang20} without losing universality.
It also has an essential connection with matrix-product states (MPS) or tensor-network states~\cite{AKLT87,PVW+07,Sch11},
which have been central in quantum many-body physics 
such as the classification of topological orders~\cite{ZCZ+15}.
Also as the analog of classical ones,
the understanding of quantum cellular automata (QCA) is greatly improved recently~\cite{CPS+17,FH19,FHH19,Arr19,Far20}.
In addition, we find the recent study of universal Hamiltonian~\cite{CMP18,KPB+20,KPB+21} implicitly relates to QCA.
There are also progress on no-go theorems regarding quantum programs~\cite{NC97,KPG19,BCA+10,YRC20}
and transversal logical gates~\cite{EK09,ZCC11,CCC+08,HNP+17,FNA+19,WA19,WZO+20,KD20,ZLJ20,YMR+20,WWC+21} by using approximations,
which broaden our understandings of universality and fault-tolerance.


Now we provide a more detailed account of this work.
We provide a systematic study of some UQCM and relations with quantum codes and quantum algorithms.
A computing model is an information processing scheme. 
Information is carried by a target set which is acted upon by processing operations. 
Universality refers to the efficient realization of any processing operations.
A computing model is universal if it can achieve such a universality.
We study UQCM with a quartet framework (Fig.~\ref{fig:bibimodel}),
with a bipartite input: data register and adversary,
and a bistage (i.e. two steps of) computation: encoding and logical gates. 
The category-I models rely on the role of adversary,
while category-II models rely on the type of logical gates.

Quantum information is carried by qubits. 
Given qubits, information processing on them can be done by 
gate operations induced by external classical control (for QCM),
or internal quantum control (for QTM), or by interactions among themselves (for QCA). 
These three category-I models represent quantum states in different ways. 
For QCM, 
a state $|\psi\ket$ is represented as $|\psi\ket=U|\psi_0\ket$ for $U$ as a gate sequence and $|\psi_0\ket$ as an easy fiducial state.
For QTM, a state $|\psi\ket$ is represented as a MPS or tensor-network state.
For QCA, a state $|\psi\ket$ is represented as $|\psi\ket=e^{iH}|\psi_0\ket$ for a local Hamiltonian $H$ 
and $|\psi_0\ket$ as an easy fiducial state, and $e^{iH}$ can also be described as a sequence of matrix-product unitary (MPU) operators.

The category-II models rely on explicit encodings of qubits and the types of logical gates. 
Quantum error-correction (QEC) codes are needed to ensure fault tolerance~\cite{NC00},
and we find some UQCM are actually based on the types of logical gates on QEC codes.
A simple type is the transversal (TV) gates, the circuit depth of which is one.
Instead of being universal,
there are codes that only permit quasi-universal~\cite{WZO+20,WWC+21} TV quantum computing (TVQC).
We find multiparticle quantum walk (MPQW)~\cite{CGW13} is a model with gates of finite-depth local-unitary forms,
while topological quantum computing (TQC)~\cite{NSS+08} via non-abelian anyon braidings 
is a model with linear-depth logical gates. 
Besides, measurement-based quantum computing (MBQC)~\cite{BBD+09} and adiabatic quantum computing (AQC)~\cite{AL18} 
are examples with dynamic codes whose code space changes during a computation,
either by unitary or non-unitary external drives or control. 

The classification described above is far from complete
while it serves as a primary framework to study various models.
Additional features or restrictions, and also hybridization
can be introduced leading to new models.
(e.g., Refs.~\cite{KW97,VWC09,BFK09,Long06,Long11,CAPV13,Aar05,AA11,BF16,WY20}). 
For instance, in QCM unitary gates are often of product form and temporally ordered.
Instead of product, a duality quantum computer uses superposition of gate operations,
inspired by the wave-particle duality principle~\cite{Long06,Long11},
which has led to quantum algorithms based on 
linear combination of unitary gates~\cite{WL16,SLL19,Zheng21} (also see Sec.~\ref{subsec:qsvt}).
From the channel-state duality~\cite{Cho75}, 
gates can also be viewed as states hence acted upon by the so-called superchannels~\cite{CDP08a,CDP08,CDP09},  
which has led to a quantum computing model without causal order~\cite{CAPV13}.
They also require a quantum system as an adversary, 
which is the analog of the machine memory in a QTM. 
Besides, there are also randomized (or probabilistic) computing,
nondeterministic computing, computing models with oracles, random-access machine, 
and complexity classes besides bounded-error quantum polynomial time (BQP) etc,
which are all important subjects in computation theory yet will not be studied here~\cite{Sav98,AB09,Gru99}.

Our systematic study turns out to be fruitful.
By classifying two categories,
we can consider the combination of models from category-I and category-II. 
This leads to a few schemes that have not been well studied. 
Finding a code that can be universal with a fixed type of logical gates could be difficult,
instead hybrid schemes that use different types of gates or operations on codes can be employed.
As a code is often carried by a quantum many-body system,
our study also highlights the central roles of entanglement, 
especially the one that is generic across a phase of matter,
as demonstrated by the recent progress of MBQC~\cite{SWP+17}.

Besides, we carry out a brief study of quantum algorithms
which are the main tools to reveal the power of quantum computing. 
Instead of studying specific problems,
here we focus on the meta-algorithm structure of quantum algorithms.
Depending on how a problem is given as input,
we study how a quantum algorithm is designed.
There are at least three types of presentation of problem: 
as bits, as qubits, or as quantum oracle operators. 
The first type is common but may not be efficient.
The second type relates to stored-program quantum computing (SPQC)~\cite{NC97},
and we provide a detailed study of it.
The third type relates to query algorithms, in particular,
we study QSVT~\cite{GSLW19} and extend it in the framework of quantum comb~\cite{CDP08a,CDP08,CDP09},
hence motivating new quantum algorithms or protocols.

This work contains the following parts.
In Section~\ref{sec:pre} we provide the basics of quantum operations,
matrix-product states, and quantum codes. 
In Section~\ref{sec:catI} we study the quartet framework of UQCM,
the concept of universality,
and the three category-I UQCM based on the universality in the space of unitary gates,
matrix-product states, and Hamiltonian, respectively.
In Section~\ref{sec:catII} we introduce category-II UQCM starting from a physical  
perspective of logical gates based on the uncertainty principle.
In particular, we analyze MBQC and MPQW from the angle of quantum codes. 
In Section~\ref{sec:hy_comb} we study hybridization and combinatorics of UQCM,
and discuss a few schemes that can be further investigated.
In Section~\ref{sec:qalg} we study the structure of quantum algorithms.
We conclude in Section~\ref{sec:conc} with perspectives on subjects that are not fully studied in this work.

\section{Preliminary}
\label{sec:pre}

Here we introduce basic notions in this work and fix notations. 
They are mainly on three subjects: quantum operations, quantum codes, and matrix-product states.

\subsection{Quantum operations}
\label{subsec:qo}

In this work, we always consider finite-dimensional Hilbert spaces. 
The pure state of a certain degree of freedom of a quantum system is a 
normalized vector in a Hilbert space, $|\psi\ket\in \C H$.
Its evolution can be unitary from a unitary group, $U\in SU(d)$, with $d=\text{dim}(\C H)$,
and $U^\dagger U=UU^\dagger=\I$,
or in general described by a quantum channel $\C E$, 
which is a completely-positive trace-preserving (CPTP) map of the form 
\be \C E(\rho)=\sum_i K_i \rho K_i^\dagger, \ee
for $\rho \in \C D(\C H)$ as generic mixed states, 
and $\sum_i K_i^\dagger K_i=\I$ as the trace-preserving condition~\cite{NC00}.  
A common class of evolution is dynamical semigroups continuous in time,
and the unitary case can be written as $U=e^{-itH}$ for a Hamiltonian $H$ and time $t$. 
We often consider local $H$ when a locality in the real space can be defined. 

\begin{figure}[t!]
    \centering
    \includegraphics[width=.1\textwidth]{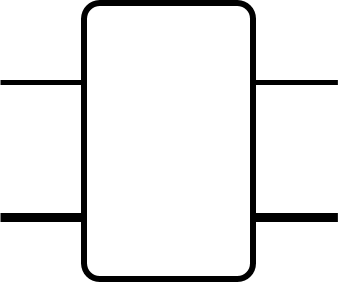}
    \hspace{1cm}
    \includegraphics[width=.35\textwidth]{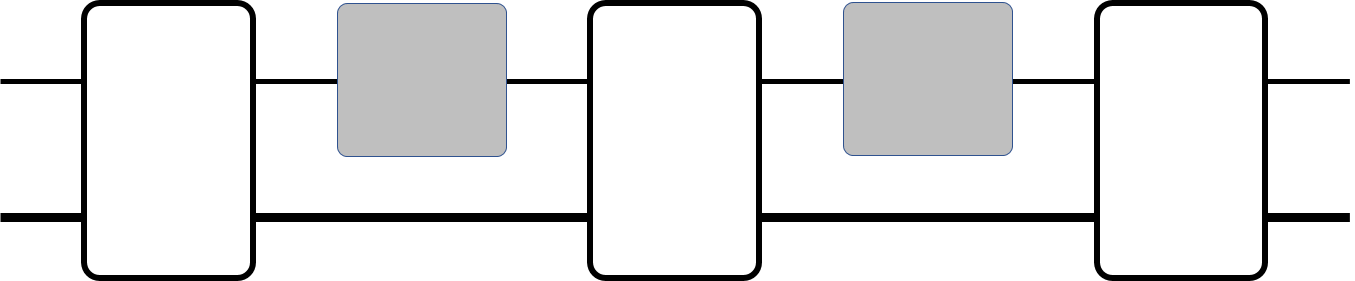}
    \caption{The quantum circuit for a quantum channel (left): the box is the unitary dilation of a channel,
    and a quantum comb (right): the boxes are unitary operators, the gray boxes are input of it.
    The top (bottom) wires carry the data (adversary).
    }
    \label{fig:qcomb}
\end{figure}

An isometry $V$ is a channel with $V^\dagger V=\I$.
A channel can also be described as an isometry with $V=\sum_i |i\ket K_i$,
and further dilated to a unitary $U$ with $V=U|0\ket$ as the first block column of $U$,
and $|0\ket$ as the initial ancillary state, see Fig.~\ref{fig:qcomb}. 
A quantum instrument is a set of CP maps $\{\Phi_i\}_{i\in \C I}$ with an index set $\C I$
and the sum of $\Phi_i$ is TP. 
It can be used in the study of measurement and error correction.

From channel-state duality~\cite{Cho75}, 
a channel $\C E$ can also be represented as a quantum state 
\be \omega_{\C E} = (\C E \otimes \I) (|\omega\ket \bra \omega|) \ee
for $|\omega\ket \in\C H \otimes \C H$ as the maximally entangled state defined on two copies of the input space. 
Such states are called Choi states in this work. 
Superchannels~\cite{CDP08a,CDP08,CDP09} can also be introduced that maps between Choi states. 
A set of channels $\{\C E_i\}$ can be taken as the input to a so-called quantum comb,
which can yield another channel or comb as the output, see Fig.~\ref{fig:qcomb}.
We find supermaps are useful to understand operations on quantum codes, 
quantum meta-algorithm, and QTM, etc.
We show that supermaps or higher-order quantum operations do not change universality, 
but they can motivate new algorithms or protocols for quantum tasks.

\subsection{Matrix-product states}
\label{subsec:mps}

\begin{figure}[b!]
    \centering
    \includegraphics[width=.3\textwidth]{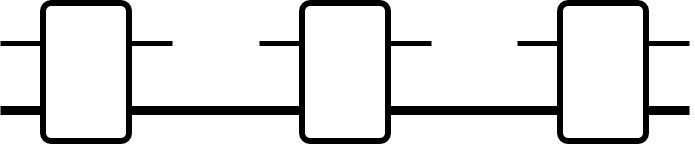}
    \caption{Quantum circuit to prepare a matrix-product state. 
    The bottom wire is the adversary, which serves as the machine memory in a QTM.}
    \label{fig:mpscic}
\end{figure}

Quantum information is often carried by pure quantum states. 
A universal state form has been developed which is known as 
matrix-product states (MPS)~\cite{AKLT87,PVW+07,Sch11}.
An MPS of a system of size $N$ takes the form 
\be |\psi\ket= \sum_{i_1,\dots,i_N} \T tr (B A^{i_N} \cdots A^{i_1} ) |i_1\dots i_N\rangle,\ee
for a boundary operator $B$.
The set of $\{A^{i_n}\}$ matrices at each site $n\in \{1,2,\dots,N\}$ form a channel $\mathcal{E}_n$
acting on a space.
We name this as the adversary space, which in other contexts may be known as
the virtual space, bond space, memory, or correlator.
The dimension of the adversary is often known as the bond dimension, denoted by $\chi$,
and $i_n\in \{1,2,\dots,\chi\}$.

We distinguish between ancilla and adversary to highlight the role of the later. 
Additional qubits besides the data qubits are often called ``ancilla''. 
The ancilla may also be part of the final state containing the solution. 
The adversary is also an ancillary system but it plays a special role,
e.g., it may interact with each data qubit.
The adversary may be carried by a distinct type of system from the data system.

An important feature of MPS is its duality: the property of the state
is encoded in the adversary.
This actually forms the foundation for a local model of quantum Turing machine~\cite{Wang20}.
An MPS can be prepared by a sequential quantum circuit, see Fig.~\ref{fig:mpscic},
and data qubits are entangled together indirectly via the adversary.
A lattice or graph can be introduced to encode the locations of data qubits,
and this extends MPS to tensor-network states with multipartite adversary.
The MPS and tensor-network states also play central roles in topological order 
and quantum many-body states~\cite{ZCZ+15}.

\subsection{Quantum codes}
\label{subsec:code}

\begin{figure}
    \centering
    \includegraphics[width=.25\textwidth]{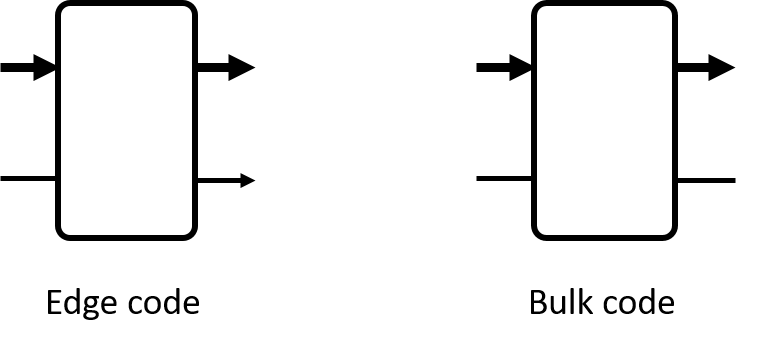}
    \caption{The edge code and bulk code.
    The arrows emphasize the input and output:
    The edge code outputs both the top wire and the bottom wire $\C H_L$, forming $\C H_P$, while the bulk code only outputs the top wire carrying $\C H_P$.
    }
    \label{fig:ebcode}
\end{figure}

\begin{figure}[b!]
    \centering
    \includegraphics[width=.4\textwidth]{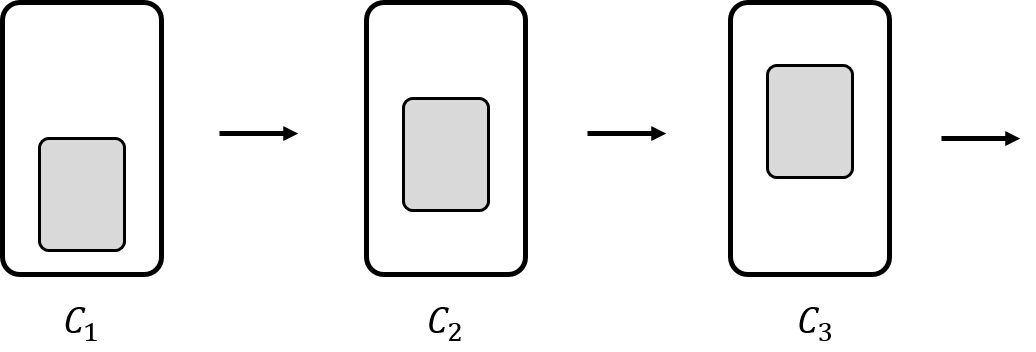}
    \caption{Schematics to show a dynamic code.
    The arrows stand for computation steps, 
    the large boxes are the whole space,
    while the gray boxes are the dynamic code spaces.}
    \label{fig:dycode}
\end{figure}

A quantum code is a subspace $\C C \subset \C H$ of a Hilbert space $\C H$.
A quantum error-correction (QEC) code is a code that can correct a set of errors. 
A set of error operators $\{E_i\}$, describing a noise model $\C N$, on a code is exactly correctable iff
\be P E_i^\dagger E_j P= a_{ij} P \ee
for $[a_{ij}]$ as a non-negative matrix and $P$ as the projector onto the code space~\cite{NC00,KL97}.
The correction scheme is defined as a recovery channel $\C R=\{R_s\}$ 
with $R_s=PF_s^\dagger/\sqrt{d_s}$ for $ P F_s^\dagger F_t P= d_s \delta_{st} P$
for $[d_s]$ as the diagonalization of the matrix $[a_{ij}]$.

This has been extended to approximate QEC of a channel $\C N$ by an optimal recovery $\C R$ on a code $\C P$ so that 
\be D(\C R \C N \C P, \I)\leq \epsilon \ee
for $\epsilon$ as the accuracy of the QEC given by a distance $D$, say, the diamond-norm distance~\cite{BO10}.
For reliable computation, $\epsilon$ shall be as small as possible.
A concrete scheme is to introduce external control parameters $\vec{\lambda}$ so that 
the function $\epsilon(\vec{\lambda})$ can be tuned to zero.
Such codes are called quasi codes and computation with them can achieve quasi universality~\cite{WZO+20,WWC+21}.

Besides the distinction between approximate codes and exact codes above, 
there are many ways to classify quantum codes.
Here we mention a few binary classifiers that will be used in this work.
Let $\C H_L$ ($\C H_P$) be the space for the logical (physical) states before (after) the encoding. 
An encoding $V:\C H_L \ra \C H_P$ defines an edge code if $\C H_P$ contains $\C H_L$,
otherwise it defines a bulk code, see Fig.~\ref{fig:ebcode}. 
A bulk code is often described by a code projector $P=VV^\dagger$ instead of the encoding $V$.
A code is local if the encoding can be defined from a set of local operators,
given a locality structure on the physical system.
A code is a stabilizer code if its codewords are defined by a stabilizer group, 
generated by a set of commuting stabilizers~\cite{Got98}.
An encoding is collective if it encodes many qubits altogether into $\C H_P$.
An exact quantum code with $n$ physical qubits and $k$ logical qubits 
is usually labelled as $[[n,k,d]]$, with distance $d=2t+1$, 
for $t$ as the maximal number of physical qubits that can be affected by noises.
A code is dynamic if $P$ changes with time due to external or intrinsic drives, 
otherwise it is static, see Fig.~\ref{fig:dycode}. 
A code is a program code if it encodes operations on a Hilbert space, 
which are from a subspace of a logical space $\C H_L$ into $\C H_P$,
otherwise it is a usual (state) code that encodes logical states into physical states. 
Some of these notions will be elaborated in the following sections.

Now we study operations on quantum codes.
Given a code projector $P$, unitary logical gates are unitary with $[U,P]=0$.
This can be extended to any CP map $\Phi$ serving as logical operations with 
$\Phi^\dagger(P)=P$, and $\C P \Phi \C P=\Phi \C P$~\cite{WWC+21}.
Other operations will make errors.
Furthermore, we can consider supermaps that map encoding to encoding operations. 
Here we mention a few basic types. 
Given two codes $C_1$ and $C_2$,
a code augmentation leads to $C_1 \oplus C_2$, and one of them, say $C_2$, is used as resources 
to realize the additional logical gates that a code $C_1$ cannot realize.
The magic-state injection~\cite{BK05,Kni05} is such an example. 
A concatenation of an encoding $V_1$ with another one $V_2$ will map $V_1$ to $V_2'V_1$ 
for $V_2'=V_2^{\otimes m}$ as a certain number $m$ of tensor-product of $V_2$.
Various tensor products of codes can be defined leading to good LDPC codes~\cite{BE21}.
There are also dynamical ways to change codes, e.g.,
by measurements or external drives.
Examples include MBQC, AQC, gauge-fixing~\cite{PR13,AD14,Bom15}, 
and switching between stabilizer codes~\cite{HCKH15}. 

\section{Universal quantum computing models: Category-I}
\label{sec:catI}

\subsection{The quartet framework}

\begin{figure}[t!]
    \centering
    \includegraphics[width=.3\textwidth]{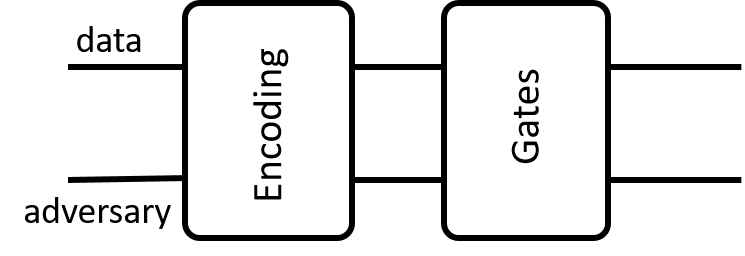}
    \caption{The quartet framework.}
    \label{fig:bibimodel}
\end{figure}

The study of models of classical computation~\cite{Sav98,AB09} is a mature field.
However, due to the weirdness and hardness of quantum information,
many puzzles and open problems are still under tense study,
and this will affect our understanding of quantum computing
and the efforts to build quantum computers.

\begin{figure}[b!]
    \centering
    \includegraphics[width=.12\textwidth]{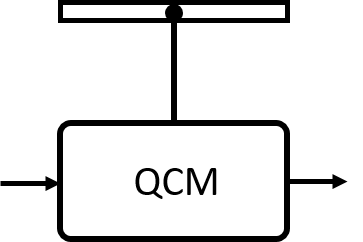}\hspace{1cm}
    \includegraphics[width=.15\textwidth]{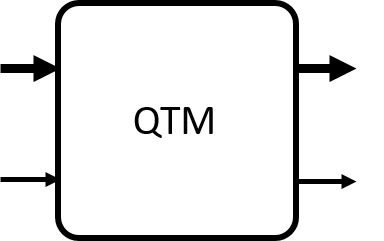}\hspace{1cm}
    \includegraphics[width=.1\textwidth]{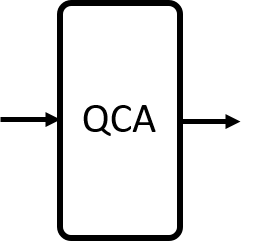}
    \caption{The three universal quantum computing models.
    QCM: the top bold wire is a classical control adversary.
    QTM: the top bold arrows are a quantum control adversary.}
    \label{fig:uqcm}
\end{figure}

In this section, we identify three primary UQCM as their classical analogs, 
via a quartet framework, Fig.~\ref{fig:bibimodel}. 
Here we assume perfect information protection, i.e., the identity gate $\I$.
The input is bipartite: data register and adversary. 
The input structure relates to the physical mechanism of information processing. 
The adversary participates the computation in a nontrivial way. 
The encoding of qubits into QEC codes is a mechanism of information protection. 
This leads to the bistage computation: the first stage is encoding, and the second stage is computation.
The readout would be the third stage, 
but we do not consider it in this work,
which does not affect our conclusions and can be further studied separately.
The three models (Tab.~\ref{tab:uqcm} and Fig.~\ref{fig:uqcm})
are of category-I, while we also identify category-II models 
according to ways of manipulations of QEC codes in the next section.



\begin{table}[b!]
    \centering
    \caption{The three category-I UQCM.}
    \begin{tabular}{|c|c|c|c|c|} \hline
        & data &  adversary & gates & universality \\ \hline
    QCM & qubits interact with adversary & classical & small  & gate simulation \\ \hline
    QTM & qubits interact with adversary & quantum & sequential & state preparation \\  \hline
    QCA & qubits interact with each other & no & local & pattern generation \\ \hline
    \end{tabular}
    \label{tab:uqcm}
\end{table}

\subsubsection{Quantum circuit model}

Among the three models, the QCM is the most popular one for quantum computing.
Due to their equivalence, a MPS or a QCA can also be described by a circuit.
But using MPS or QCA have advantages in some proper settings.
Here we recall a few facts of QCM briefly.
The QCM is defined as the general scheme to use a gate sequence $\prod_i U_i$ acting on the initial data state $|\psi\ket$ to obtain the final state $|\psi_f\ket$ (up to measurements).
Each gate can be described classically, including its form, spacetime location in the circuit.
A gate is often realized by external means, i.e., it is not from inherent dynamics of the system.
There is no entanglement between the system and the external controls.

The essential task in QCM is to perform state conversion 
\be |\psi_f\ket= U|\psi_0\ket.\ee
Note ancilla and probability may be involved sometime, 
but here we consider the most succinct form.
A quantum gate $U\in SU(d)$ for any $d$ is often decomposed as a sequence of small gates 
\be U \approx U'= U_n U_{n-1} \cdots U_2 U_1\ee
each acting on a constant number of qubits.
The smallness of gate often can be transferred as a locality of gate for implementation.
For universality, a gate set is called a universal gate set for $SU(d)$ 
if it generates a dense subset of the group $SU(d)$~\cite{NC00,BBC+95,DN06}.
The notion of universality will be further discussed in Sec.~\ref{subsec:univse}.

\subsubsection{Quantum Turing machine}

The Turing machine is a primary computing model for classical computation~\cite{Sav98,AB09}.
QTM was established as the fully quantum version of the classical Turing machine~\cite{Deu85,BV97,Yao93}.
The read/write head, which monitors the interaction between the tape and the machine memory, 
is also quantized, and this leads to nonlocal interactions~\cite{Mye97,Oza98,Shi02,For03,MW19,Wang20,GMM20}.

A model of QTM with locality of the head is recently proposed,
and shown to be equivalent to the QCM by simulation method~\cite{Wang20}.
The locality in QTM refers to the definite location of the machine head. 
The adversary (i.e., the machine memory) can decouple from the data after the computation
\be |\psi_f\ket|a'\ket=U_\textsc{qtm}|\psi_0\ket|a\ket, \ee
achieving the state conversion task $|\psi_0\ket \mapsto |\psi_f\ket$.
Importantly, it shows that when the interaction between the adversary and data register is sequential,
a QTM prepares MPS.
It is enough to consider sequential QTM since the MPS form is universal. 
The dimension of the adversary quantifies the entanglement of the MPS. 

For a QTM, the structure of the adversary becomes essential.
MPS usually refers to cases of small bond dimensions and 1D systems. 
Higher dimensional states are usually called tensor-network states or PEPS~\cite{Sch11}.
More generally, a higher-order MPS can be introduced when the bond dimension is large
to capture the structure of the adversary.
We will describe this in Sec.~\ref{subsec:topQTM}.

\subsubsection{Quantum cellular automata}

In classical computation, CA is a computing model that arranges bits on a lattice or graph, 
and uses parallel local rules to update bits simultaneously~\cite{TM87}.
It is often used to simulate dynamics or generate patterns.
As the name indicates, a CA evolves autonomously due to local (cellular) interactions among cells.   
The autonomous evolution steps does not have to be continuous in time,
so error correction can be done between steps when necessary. 

QCA also has a rather long history in quantum computing~\cite{Arr19,Far20,Wie08}.
The time evolution can be continuous or digital,
and their universality has been established~\cite{Wat95,KW97,SW04,Rau05,Osb06,PC07,NW08,VC08,AG12}.
Different from QCM and QTM,
QCA requires a locality in the real space. 
This is an analog of quantum many-body system which also has well-defined locality in the real space. 
We follow~\cite{Osb06,PC07,NW08,VC08,RWW20} to treat QCA as the computing model relying on Hamiltonian evolution,
in particular, local Hamiltonian.



A local Hamiltonian 
\be H= \sum_{i=1}^m H_i \ee
on a system of size $n$ is usually defined as a sum of $m=\text{poly}(n)$ local terms $H_i$,
each acting on a constant number of neighboring sites.
Note here we require a geometric locality instead of a smallness of the support of $H_i$. 
A QCA is the evolution $U=e^{itH}$,
which can be digitalized using the Lie-Trotter-Suzuki operator-product formula,
and each step containing commuting local evolution is an infinitisimal QCA~\cite{PC07}.
With translational invariance, each step is globally equivalent to another step. 
A QCA changes the entanglement of the data state.
Using MPS form of data states, 1D QCA was identified with MPU operators~\cite{CPS+17}.

Furthermore, since QCA relies on the set of local Hamiltonian,
its universality can be deduced from the universality of Hamiltonians, 
which was recently established~\cite{CMP18,KPB+20,KPB+21}.
A Hamiltonian $H'$ simulates a target $H$ if its spectrum is well approximated up to an encoding.
If the error is $\epsilon$, then the time evolution $U=e^{-itH}$ will have error $\epsilon t$,
linear with time growth. 
It shows that there are universal models even in
translationally invariant spin chains in 1D.
This is consistent with the universality of 1D QCA. 
This naturally serves as a scheme for quantum simulation with parallelism.


\subsection{Universality}
\label{subsec:univse}

Universality is the central concept for UQCM. 
Here we define the notion of universality, discuss a few issues
and some of its extensions. 
A universal computing model is an information processing scheme that can achieve universality of a set of operations on a target set. 
The standard target set is a Hilbert space $\C H$ of pure states. 
The map between any two states $|\psi_0\ket \mapsto |\psi_f\ket$
is described by a unitary operator. 

The three models QCM, QTM, and QCA can achieve this universality (of state transformation) by representing states in different ways:
as a gate sequence, as a tensor network, or as a dynamical evolution.
For QCM, 
a state $|\psi\ket$ is represented as $|\psi\ket=U|\psi_0\ket$ for $U$ as a gate sequence and $|\psi_0\ket$ as an easy fiducial state.
For QTM, a state $|\psi\ket$ is represented as a MPS or tensor network state.
For QCA, a state $|\psi\ket$ is represented as $|\psi\ket=e^{iH}|\psi_0\ket$ for a local Hamiltonian $H$ 
and $|\psi_0\ket$ as an easy fiducial state, and $e^{iH}$ can also be described as a sequence of MPU.
This also extends to time-dependent $H(t)$,
such as adiabatic and Floquet ones for QCA.

The equivalence of the three models can be shown by simulation. 
Given a QTM or QCA, it can be easily simulated in QCM since a circuit can be easily constructed.
The simulation of QCM by the (local) QTM is due to~\cite{Wang20} (see also~\cite{MW19}).
As QCA can be easily described as circuits, then can be simulated by QTM.
A simulation scheme of QCM by QCA is from~\cite{Rau05} (see also~\cite{PC07,NW08,VC08}). 
As QTM can be easily described as circuits, then can be simulated by QCA.

A set of unitary gates is universal if any unitary $U\in SU(d)$ can be efficiently approximated by sequences of gates from this set to arbitrary accuracy $\epsilon$, for $d=\text{dim}(\C H)$.
For instance, the set of Hadamard H, T, and CZ gate, and also identity $\I$ is a universal gate set. 
We refer to the cost of decomposing a gate $U$ as a sequence of gates from a universal gate set $S$ as the compiling cost.
Logical gates from $S$ are usually assumed to be realized exactly.
The identity gate $\I$ from $S$ is nontrivial since it requires a perfect memory or error correction. 
We refer to the cost for achieving a perfect memory as the memory cost or coding cost. 
When $\I$ cannot be achieved, e.g. due to decoherence or approximate error correction, universality cannot be realized, strictly speaking. 
Recently, a slight variation is introduced based on quasi codes, which can be tuned to be exact codes by changing parameters, $\vec{\lambda}$. 
A quasi code can realize $\I$ with accuracy $\epsilon(\vec{\lambda})$, which can be as close as zero by changing $\vec{\lambda}$. 
The quasi universality is defined as the approaching to the (exact) universality when $\epsilon(\vec{\lambda})$ goes to zero~\cite{WZO+20,WWC+21}. 
By definition, the quasi universality is weaker than universality.

The universality above does not account for the post-processing on the output state $|\psi_f\ket$.
The post-processing can be various tasks, e.g., extraction of an operation from it, or measurements, and this is not guaranteed to be performed efficiently.
We find it is unnecessary to extend universality to account for the post-processing since some tasks cannot be done efficiently. 
Estimation of an unknown gate is such an example~\cite{CAPS04}, 
the accuracy of which is constrained by the Heisenberg limit.
There are also other types of universality referring to other target set, 
which can be a group, subspace, etc. 
This reduces to state conversion, together with post-processing, 
if the target set is embedded in a Hilbert space.
We can also consider conversion of quantum operations. 
This is described via quantum supermaps and higher-order operations~\cite{CDP08a}. 
By encoding channels as Choi states or other types of program states embedded in a Hilbert space, 
quantum superchannels can be realized by unitary operations on it.
The effect of an inefficient post-processing can be viewed as the reduction from universality to quasi universality.

How to achieve state conversion $|\psi_0\ket \mapsto |\psi_f\ket$? 
This depends on how they are described in a given problem,
leading to various types of quantum algorithms. 
If their explicit forms are known, we can find a unitary $U_0$ and $U_f$ with $U_0 |0\ket=|\psi_0\ket$
and $U_f |0\ket=|\psi_2\ket$ for a fiducial simple state $|0\ket$. 
Then $U_0^\dagger U_f$ achieves the task. 
If their explicit forms are unknown, but some oracles are given, then we need to use oracle algorithms,
such as Grover's search algorithm or SPQC. 
This will be further discussed in Sec.~\ref{sec:qalg}.


\section{Universal quantum computing models: category-II}
\label{sec:catII}

Now we study category-II models, 
which are models with explicit encodings. 
Here we briefly summarize the models we study. 
Transversal quantum computing
(TVQC) employs transversal gates that do not change entanglement of codewords,
hence the codewords contain almost all the source for universality.
Multiparticle quantum walk (MPQW) uses interactions among excitations, 
which induce finite-depth logical gates,
and the codewords are not ground states in a static quantum phase of matter.
Topological quantum computing (TQC) employs codewords from the fusion space of anyons, which shall be far apart to avoid interactions,
and non-abelian anyon braidings 
can change entanglement but not the topological (TOP) order (``long-range entanglement'').
Adiabatic quantum computing (AQC) employs continuous-time code `drift' which also changes entanglement.
Measurement-based quantum computing (MBQC) and some code switching schemes 
employ code switching that changes entanglement,
hence universality relies on all the codes appeared under switching.
The above models can use ground states, edge modes, excitations, or defects from quantum phases of matter as the codes. 


\subsection{Types of logical gates: uncertainty principle}

Given a code, we often consider unitary logical gates with $[U,P]=0$.
The operation $U$ will take time, so during its implementation it may lead to states outside of the code space. 
The depth of the operation $U$ could be larger than one,
so it is crucial to make sure it can be done fault-tolerantly. 
Besides, measurements are also commonly used to yield logical gates.  
It is also crucial to make sure measurement errors can be corrected, e.g., by repetitions.
One often only considers simple projective measurements since measurements described by POVM can be reduced to projective ones on a larger space. 

A logical gate can be described as a circuit, up to projective measurements. 
Its complexity can be measured by the features of the circuit, 
such as its size (number of gates), depth, locality, 
or other quantities
up to non-entangling swap or permutation gates.
Among them, the depth of local circuits is an important feature.

We find that the uncertainty principle has a relation with the depth of logical gates.
In Heisenberg picture, the depth can be viewed as the maximal evolution time of local operators 
under a circuit. 
Given a pure state $|\psi\ket$ and two non-commuting operators $A$ and $B$, 
the standard Robinson-Schr\"{o}dinger uncertainty relation is 
\be (\Delta A)^2 (\Delta B)^2 \geq | \frac{1}{2} \bra \{ A,B\} \ket - \bra A \ket \bra B \ket |^2 
+|\frac{1}{2i} \bra [A,B] \ket |^2 \ee 
for the standard deviation $\Delta A:= \sqrt{\bra A^2 \ket-\bra A \ket^2}$
with $\bra A \ket:= \bra \psi| A |\psi \ket$.
Now choose $B$ as a Hamiltonian $H$, which could be time-dependent,
and define $\tau_A:=\Delta A / \bra \dot{A} \ket$ as a time measure for $\dot{A} = \frac{d}{dt}A$,
and $\bra [A, H] \ket \neq 0$, then 
\be \tau_A \Delta H \geq \frac{1}{2}, \ee
as the time-energy uncertainty relation.
We can define $\Delta \tau:= \min_A \tau_A$ as the minimal time.

For a codeword state $|\psi\ket$, unitary logical gates change the value of local observable in Heisenberg picture.
When all logical gates are of depth one, this means the energy uncertainty $\Delta H$ on the code space diverges. 
This applies to TVQC.
When logical gates are of finite depths, 
product of such gates can increase the depth. 
The length of such product in general is independent of the physical system, 
so there is no definite behavior of $\Delta H$.
MPQW is such an example which uses interactions among excitations. 
But if the length of a gate sequence is proportional to the system size,
the uncertainty $\Delta H$ shall be suppressed by increasing the system size.
When all logical gates have diverging depths,
it means the code space has to be degenerate.  
This applies to non-abelian anyon braidings for TQC. 
For AQC with a unique ground state, the time-energy uncertainty relation shows that
the total adiabatic evolution time $t_f \sim O(h/\Delta^2)$ for $h:=\max_s \|\dot{H}(s)\|$,
and $H(s)$ ($s\in [0,1]$) as the Hamiltonian,  
$\Delta$ as the minimum of the gap above the ground states~\cite{FG96}.
The dependence of $\Delta^2$ instead of $\Delta$ is only apparent since $\|\dot{H}(s)\|$ 
contains an energy scale. 
For code switching schemes, logical gates change stabilizers (or  check operators).
A switching is often of finite depth, and this leads to no explicit constraints on $\Delta H$.
Below we survey each model by specifying its definition, application, exemplary codes and QEC property. 

\subsection{TVQC and related}
\label{subsec:tvqc}

Given a QEC code, probably the simplest type of unitary logical gates is transversal (TV) gates. 
Given a block partition $\C J=\{j\}$, which is a set of labels $j$ for subsystems that allow error correction on it,
a TV logical gate $U$ takes the form 
\be U=\otimes_j U_j, \ee 
which do not spread out local independent errors on each block $j$, 
and also do not change entanglement with respect to the block partition. 
It is well known that given a QEC code the number of TV logical gates on it is a finite number,
hence not universal~\cite{EK09,ZCC11,CCC+08}. 

In fact, TV schemes have a longer history in quantum metrology. 
The central task is to estimate unknown parameters or operations, and the estimation accuracy is limited. 
With $n$ queries to the unknown object,
using product states can obtain the shot-noise scaling $1/n$,
while using optimal entangled states can obtain the Heisenberg scaling $1/n^2$ for the standard deviation~\cite{TA14}. 
There are so-called super-Heisenberg scalings, but a careful analysis in terms of query complexity to the unknown objects shows the optimality of the Heisenberg limit~\cite{ZDK10}. 

The model of universal SPQC (see Sec.~\ref{subsec:spqc}) can be treated as special type of TVQC. 
Recall that a universal SPQC is to realize any $U$ by $G|\varphi\ket |P_U\ket= U|\varphi\ket |P_U'\ket$ for program state $|P_U\ket$ and data state $|\varphi\ket$, and a universal unitary operation $G$ that is independent of the program and data.
We can define an encoding $V:|f\ket \mapsto |\varphi\ket |P_U\ket$
for $|f\ket=U|\varphi\ket$. 
This encoding $V$ is covariant. 
As has been proved~\cite{BCA+10,YRC20}, the accuracy of SPQC is constrained by the Heisenberg limit. 
Note that this does not apply to SPQC with non-universal programs, such as magic-state injection.

Transversality for a set of gates can be achieved by requiring the covariance of encoding,
a.k.a. covariant codes.
Recently Lie-group covariant codes are studied and limitation on accuracy is proved~\cite{HNP+17,FNA+19,WA19,WZO+20,KD20,ZLJ20,YMR+20,WWC+21}.
The relation with metrology is recognized~\cite{KD20,ZLJ20,YMR+20}.
For $SU(d)$ covariant codes, the $SU(d)$ group of gates can be done but QEC cannot be done exactly. 
Instead of universality, only weaker versions can be achieved such as the quasi universality~\cite{WZO+20,WWC+21}.



\subsection{MPQW and FDLU gates}
\label{subsec:mpqw}

Quantum walk has been essential to design fast quantum algorithms.
There are many models or algorithms in the name of quantum walk,
recently the idea of using reflection or oracle leads to the QSVT algorithm~\cite{GSLW19}.
Compared with its relevance with algorithms, 
the connection with quantum code is less studied.
We show that MPQW is an example of category-II UQCM 
and logical gates by finite-depth local unitary (FDLU),
and also with features of QCA. 
This motivates variations of MPQW in the light of QCM.


We analyze the MPQW based on Bose-Hubbard model~\cite{CGW13},
and our study also applies to other relevant models~\cite{Jan07,Nag10,Nag12,BHS+15,LT16,TGL+16}.
Consider a system on a lattice (or graph),
with each lattice site carrying a boson, fermion, or qudit (as distinguishable particle),
and time-independent local Hamiltonian $H$ defined on the lattice specifies the local interactions among the particles.
Qubits are carried by excitation wavepackets with particular momentums $k$.
We assume the initial state is a product state, 
and it evolves under $H$ to the state at time $t$ as
\be |\psi(t)\ket= e^{-itH} |\psi(k_1,k_2,\dots,k_n)\ket.\ee
No external control is required during the evolution.
To simulate a quantum circuit, special lattice structure is needed. 
Each gate corresponds to a special local region on the lattice.
The locality is crucial for initialization, gates, and also measurements.

Treating the whole system as a code,
it is not the ground states of many-body phases of matter.
It employs a large section of the spectrum of a model. 
It is not an exact code and its exact code distance is small.
The wavepackets cannot be ideally prepared, and its finite width causes errors~\cite{CGW13,BHS+15,LT16,TGL+16}.
As wavepackets are not infinitely far apart, thermal excitations can cause small disturbance to the data qubits,
hence small logical errors. 
The local excitations can be detected by measuring local Hamiltonian terms 
and corrected approximately by cooling them locally. 
Depending on the form of the wavepackets, FDLU gates are approximate instead of being exact.
Besides, a finite lifetime of wavepackets is also a source of decoherence.
There could also be other systematic errors, e.g.,
static disorder of $H$ may lead to localization~\cite{LT16}.

To improve its fault-tolerance, one can consider code concatenation or use other codes. 
Besides Hubbard model, one can employ other quantum many-body systems with local Hamiltonian,
such as topological Mott insulators or spin chains which could have nontrivial symmetry or topology protection,
and using excitations and their scatterings for gates. 
The MPQW is also autonomous, so it has some features of QCA. 
A spatial direction is chosen as the simulated time. 
Just as photons, the injected wavepackets will evolve until they are being measured. 
However, a UQCM with FDLU gates does not have to be autonomous. 
One can in principle use controls to enact a FDLU gate in the spirit of QCM.
A proposal will be discussed in Sec.~\ref{subsec:fdluc}.




\subsection{TQC and HDLU gates}
\label{subsec:tqc}

Logical gates can also be of high-depth local unitary (HDLU) forms.
Here, being ``high'' means the depth is not a constant independent of the code distance.
Due to the high depth, a HDLU gate can spread out local errors,
so QEC is needed during its implementation to prevent this.

A notable example is TQC based on non-abelian anyons~\cite{NSS+08}.
TQC has been well developed and it also has algorithmic applications,
such as to compute Jones polynomials~\cite{FKW02a}.
Currently, TQC is the highly expected candidate to build fault-tolerant quantum computers in the future. 
Due to its maturity, here we only recall some facts briefly. 

A braiding is typically of linear depth~\cite{KKR10,BD12,ZHB17}.
This is due to the topological degeneracy of the fusion space.
Although braiding does not need to be adiabatic,
it is quasi-adiabatic in the sense that it needs to be much slower than 
the time scale set by the gap $\Delta$,
but fast enough to avoid dynamical effect due to the quasi-degeneracy of the fusion space~\cite{HW05}.
Although braiding is a 2D phenomenon, 
it can be realized in network of 1D wires, 
such as Majorana zero modes, which act as Ising anyons~\cite{SFN15}. 
Anyons are naturally hosted by topological systems as local excitations,
and their classifications are rapidly developed in recent years~\cite{ZCZ+15}.
A sequence of gates is simulated by a sequence of braidings,
which indeed change the local details of the system but not the topological order. 
The exact code distance of a topological code is large,
which is a major advantage of the class of topological codes.


\subsection{AQC and dynamic codes}
\label{subsec:aqc}

AQC is an example of QC with dynamic code.
We first recall AQC and then generalize to dynamic code.
AQC has been applied in many quantum algorithms, such as quantum annealing,
adiabatic formulation of Grover's search, 
optimization problems, etc~\cite{AL18}.
Its universality was proved based on the Feynman-Kitaev circuit-to-Hamiltonian (FKCH) map~\cite{KSV02}.
Namely, given a gate sequence $U=U_N\cdots U_2 U_1$ and initial state $|l\ket$,
the FKCH map defines 
\be H_\text{FKCH}=  - \sum_{t=1}^N ( U_t\otimes|t\ket \bra t-1|+ h.c.) 
+ H_\text{clock} + H_\text{edge}  \ee
which takes the history state
\be |\psi_l\ket := \frac{1}{\sqrt{N+1}} \sum_{t=0}^N V_t |l\ket |t\ket \ee 
as a ground state for $V_t:=U_t\cdots U_2 U_1$ and $V_0=\I$,
$\{|t\ket\}$ as states of a clock, and $H_\text{clock}=\sum_{t=0}^N |t\ket \bra t|$,
$H_\text{edge}$ specifies the initial and final states. 
The basic method is to employ a Hamiltonian $H(s)=(1-s)H_0+s H_f$ for $s\in [0,1]$ as an effective adiabatic parameter, 
and the final state is contained in the ground state of $H(1)$.
The model $H(s)$ at each instantaneous time $s$ may not be exactly solvable but remain gapped. 
Instead of gate simulation, the goal of AQC here is to prepare states.
It is proved that 1D system (with 9-state particles) can be universal for AQC~\cite{AGIK09}.

The input state $|l\ket$ can be a codeword from a QEC code. 
Besides thermal noises, the main source of errors is the adiabatic passage,
which could suffer from diabatic errors~\cite{AL15,YSK13,GRK+19}.
The total adiabatic evolution time $t_f \sim O(h/\Delta^2)$ 
is always finite, 
for $h:=\max_s \|\dot{H}(s)\|$,
$\Delta$ as the minimum of the gap above the ground states~\cite{FG96}.

The model $H_\text{FKCH}$ does not correspond to a definite type of phase of matter.
The clock can be replaced by qubits in order to define a better notion of locality.
We can directly use ground states of phases of matter for AQC. 
If qubits are encoded in the universal features of a phase,
then the adiabatic evolution does not make any logical change.
When a gapped passage between two phases of matter exists, 
AQC can realize the state conversion from one phase to the other. 
Also note instead of adiabaticity, 
quasi or local adiabaticity is defined for phases of matter~\cite{HW05,CGW10,SM16,BRF17}.

For dynamic code, the code $P$ does not come back to itself after the external drive, $U$. 
Instead, it is changed to another code with projector $P'=UPU^\dagger$. 
Also the encoding isometry $V$ is changed from $V$ to $V_f=UV$. 
However, this will lead to $U$ as merely updating the encoding without any logical operations. 
To avoid this, we need to use a fixed encoding $V_*$ according to some scheme.
For instance, if codewords have different energies, 
then the codewords at each stage can be ordered energetically.
Say, for $C$ a basis of codewords $\{|\lambda_1\ket,|\lambda_2\ket,\dots\}$ 
and for $C'$ a basis of codewords $\{|\lambda_1'\ket,|\lambda_2'\ket,\dots\}$ both with increasing energies, 
and an encoding $V_*$ can be defined by mapping $|\lambda_i'\ket\mapsto |\lambda_i\ket$. 
QEC needs to be defined relative to each projector at each stage. 
This applies to AQC which usually only has one codeword as the ground (or excited) state of a Hamiltonian,
and also applies to code switching schemes by nonunitary operations such as measurements.



\subsection{MBQC and code switching}
\label{subsec:mbqc}

In MBQC, gates are induced by local projective measurements (PVM) on a resource state, 
whose outcomes are feed forward to later ones. 
MBQC has many applications, such as for graph states, 
blind quantum computing, linear optics, etc. 
In this section, we treat MBQC as a code switching scheme on some edge codes;
alternatively, our work can be viewed as an extension of MBQC via code switching.


A 2D cluster state on the square lattice is
\be |C\ket = \prod_{\bra i,j\ket} CZ_{\bra i,j\ket} |+\ket^{\otimes n} \ee
with CZ gate applied on the two qubits on each edge $\bra i,j\ket$.
To simulate a circuit, a cluster direction is set as the simulated time,
and a logical qubit is assigned to a wire,
with junctions connecting wires to simulate CZ gates.
A site on the cluster is deleted by a Z-basis measurement.
Now motivated by the code switching idea, 
we can use a ``two-way'' scheme with the turning on and off the CZ gates
and refreshment of qubits.
It is easy to see two qubits are enough to realize any qubit gate by teleporting back and forth,
and four qubits are needed to realize a CZ gate. 
The qubits can be arranged in 1D, 
and there is no direction for the simulated time. 
However, the one-way scheme with 2D cluster state
and others~\cite{RB01,BBD+09,Wei18,Wang19b} 
has many advantages as the resource state can be given offline. 



To increase the fault-tolerance or code distance, 
other resource states have been studied.
Recently, MBQC has been related to symmetry-protected topological (SPT) phases of matter~\cite{SWP+17,ROW+19}.
However, the distance is small due to the short-range entanglement so that information is localized.
Anyon braidings simulated by measurements have been developed for toric codes~\cite{RHG06,FMM+12}
and in the measurement-only TQC~\cite{BFN08,ZDJ16,BF16}.
This motivates the usage of symmetry-enriched topological (SET) ordered states~\cite{MR13,WBV17,BBCW19}, which are both SPT and TOP ordered.
Here we sketch the basic ideas while details are needed for separate further study.
For a SET order with $G_1$ SPT order and $G_2$ TOP order,
different logical gates can be deduced from them.
Anyons are nontrivial representations of the symmetry $G_1$. 
Using MPS and PEPS theory, 
a resource state in a phase can be represented as an on-site local tensor $\{A^i\}$ of the form
\be A^i=\bigoplus_{nm} |n\ket\bra m|\otimes B^i_{nm} \otimes C^i_{nm} \ee
for each site $i$
and proper boundary conditions for initial logical states. 
The block structure relates to the TOP order, 
the $C$ part is due to the symmetry $G_1$, while the $B$ part is for the junk details of each state.
The logical states live in the first part, 
while now the symmetry $G_1$ can have a nontrivial action on it. 
Using the encoding of holes, TOP logical gates can be done by braiding of holes, 
while SPT logical gates can be done by PVM on the boundary of holes, 
which change the size of them.
The holes need to be separated far apart enough to account for the size changes to maintain a large code distance.



The one-way MBQC deletes stabilizers or check operators for a code.
In general, code switching between two codes $C_1$ and $C_2$ could involve fault-tolerant preparation of new check operators. 
The gauge-fixing~\cite{PR13,AD14,Bom15} is a type of code switching on subsystem codes which have gauge qubits. 
For stabilizer subsystem code $C_1$ with gauge group $G_1$ and stabilizer group $S_1$, and similarly for $C_2$, 
then $C_2$ can be gauge-fixed from $C_1$ when $S_1\subset S_2 \subset G_2 \subset G_1$~\cite{Bom15}.
This idea can be applied to combine TV gates from two codes. 
Starting from the state $|\psi\ket \in C_2$, applying a TV gate $U$ natural for $C_1$ but not for $C_2$ leads to $U|\psi\ket \in C_1$. 
Then gauge-fixing the state can drive back in $C_2$ with a logical action $U'$ on $|\psi\ket$.

\section{Hybridization and Combinatorics}
\label{sec:hy_comb}

\begin{table}[]
    \centering
    \begin{tabular}{||c|c|c|c|c|c||}\hline \Xhline{3\arrayrulewidth}
        & TVQC & MPQW & TQC & AQC & MBQC        \\ \hline
    QCM &  \cellcolor{gray} I  &  \cellcolor{gray} II  & $\checkmark$  &  $\checkmark$  &  $\checkmark$        \\ \hline
    QTM &  $\checkmark$  &  $\checkmark$  & \cellcolor{gray} III  &  $\checkmark$  &  $\checkmark$        \\ \hline
    QCA &  $\checkmark$  &  $\checkmark$  & $\checkmark$  &  \cellcolor{gray} IV  &  $\checkmark$        \\ \hline \Xhline{3\arrayrulewidth}
    \end{tabular}
    \caption{Combinatorics of UQCM. 
    The $\checkmark$ marked cells refer to schemes that have been studied.
    The numbered cells refer to the relatively new schemes that we identify.}
    \label{tab:hym}
\end{table}

In this section, we put the two categories of UQCM together and study their hybridization and combinatorics.
Here hybridization means a mixing within a category,
e.g., using different types of logical gates, or using hybrid adversaries.
When a ``pure'' universal scheme proves to be challenging,
hybrid schemes can be employed. 
Hybridization provides a large room for fault-tolerant universal QC.
There are many hybrid schemes in literature, 
such as magic-state injection~\cite{BK05,Kni05},
color codes using topological TV gates and code switching~\cite{Bom15}, 
coupled sine-Gordon qubits using topological TV gates and code drift~\cite{Wang20a}.

Below we focus on combinatorics of UQCM, 
which refers to different ways to combination information processing and information protection.
See Table~\ref{tab:hym}.
By putting the two categories of models on different dimensions of the table, 
we identify a few promising schemes that have not been well studied,
to the best of our knowledge. 
In order to clarify the meaning of combinatorics of models,
we first survey the schemes that have been studied in literature, summarized in Table~\ref{tab:combuqcm}. 
Note the schemes shown here are only examples,
there could be diverse tasks that can be defined and studied.
We label each scheme by its column and row names.



Given the diversity of quantum computing schemes,
it is straightforward to compare them but difficult to tell which one is better.
There are trade-offs on properties of a scheme,
such as the code distance, encoding rate, type of gates, accuracy, overhead, 
and the ease of experimental implementation, etc. 
Below we study four schemes that have relatively not been explored in depth,
and each scheme leads to different request on states or dynamics.

\subsection{Scheme I: Approximately covariant codes} 

Using TV operations to realize a sequence of gates is probably the simplest quantum computing scheme.
We mentioned that 
$SU(d)$ covariant codes are constructed which is transversal for $SU(d)$ by definition, 
yet as QEC is not exact only quasi universality can be achieved.
Besides, the allowed gates are not digital in the sense that 
there is no need to decompose a target gate $U\in SU(d)$ into a sequence of primary gates.

What have not been explored are codes that are approximately covariant,
especially codes that allow a discrete set of TV gates. 
For instance, the gate set of H and T gates. 
It turns out, if there were, such codes are unusual. 
As both H and T are TV operations, the product of them are also TV.
This set approximates the group $SU(2)$ by a product of H and T.
Such a code would have $SU(2)$ as an approximate symmetry instead of exact symmetry. 
Even $U(1)$ can only be approximate since product of H and T cannot yield exactly diagonal gates~\cite{Sel12,KMM13}.
This requires that there is no infinitesimal TV logical gates,
otherwise, the whole group $SU(2)$ is TV.

Denote an infinitesimal logical gate as $\I_\epsilon$ which is $\epsilon$-close to $\I$ gate,
and $\epsilon$ could be tunable. 
Now suppose a code does not allow $\I_\epsilon$,
then it could be a quasi-exact code~\cite{WZO+20,WWC+21} or a partially self-correcting code~\cite{BH11,BH13}.
Such codes do not allow TV $\I_\epsilon$, neither non-TV $\I_\epsilon$.
Physically, $\I_\epsilon$ should be due to local noisy processes that cannot be fully corrected.
The noisy processes do not have to be TV,
but here we assume they are. 
Therefore, neither $\I_\epsilon$ nor H and T gates can change entanglement of the code 
(e.g., the bond dimension if the codewords are MPS).

What the gates and noises can change is energy. 
The uncertainty principle implies the standard deviation $\Delta H$ is large,
namely, the codewords cannot be degenerate.
Now we assume the code is local, i.e., defined by a local Hamiltonian $H=\sum_n H_n$,
which may be gapped or gapless.
What are the physical origins of H and T gates?
They shall relate to the symmetry of the system.
Recall that TV gates are defined up to permutation operations that only shuffle local sites.
The physics of H and T gates could be distinct. 
For instance, the toric code~\cite{Kit03} allows a TV H gate which is the global $\otimes_n \text{H}_n$ 
followed by lattice permutation, but it does not have a TV T gate.
For the family of (gauge) color codes~\cite{Bom15}, the TV H gate is from a duality of CSS codes,
and the TV T gate is from geometry of the lattice, 
and there are also TV X and Z gates. 
However, these codes are exact codes and codewords are degenerate.
The TV T gate can be obtained from a class of $SU(n)$ valence-bond solids~\cite{WAR18} (for $n=8$), 
which is a flux insertion operation.
It disturbs the Hamiltonian slightly, so it is a quasi-exact logical gate.

So far, we do not find such a code, but there are candidate systems which need separate study.
A type of gapped system is SET systems~\cite{MR13,WBV17,BBCW19} with global $SU(n)$ symmetry and abelian gauge symmetry.
A TV T gate is expected from flux insertion,
and TV H gate from duality symmetry of the system.
Gapless systems can be described by CFT~\cite{FMS97}.
It is known that monodromy of correlation functions of primary fields is equivalent to the holonomy 
of anyon braidings~\cite{Read09},
which is not TV in general.
It is not clear what could be the physical origins of H and T gates.
Besides, such a code can be a subsystem code, 
just like using the FM phase of 2D Ising system to encode a classical bit.
The energy landscape of the spectrum to encode the qubit can be complicated
so that it may take a long time to realize a transition between some pair of states.
There are systems which break ergodicity,
such as many-body localized systems~\cite{AAB+19},
spin glass systems~\cite{Nis01}, 
a notable feature of which is that it could take exponentially long for a state to relax to the ground state,
and also Floquet systems which could have approximate symmetry and take exponentially long to thermalize~\cite{HRR+20}.


\begin{table}[t!]
    \centering
    \caption{Some schemes in literature as examples of combinatorics of models.} \vspace{.1cm}
    \begin{tabular}{|p{0.13\linewidth}|p{0.8\linewidth}|}  \Xhline{3\arrayrulewidth}
      QCM-TQC   &  A well-developed scheme is to use anyon braidings to simulate gates~\cite{NSS+08}. \\ \hline
      QCM-AQC   &  The holonomic QC~\cite{PZ01} can be seen as such an example,
which uses adiabatic or non-adiabatic non-abelian geometric phases for gates.
Each gate needs a different geometric path in the control-parameter space. \\ \hline
QCM-MBQC & A well-developed scheme is to simulate gates by cutting off qubit wires and junctions from a 2D resource state~\cite{BBD+09}. \\ \hline
QTM-TVQC & A simple scheme is to use TV operations acting on the physical sites of a MPS.
The family of valence-bond solid codes is an example~\cite{Wang20a}, 
which uses a global symmetry $SU(d)$ to construct logical gates on a logical space induced by symmetry breaking. \\ \hline
QTM-MPQW & We consider FDLU circuits on MPS, which change its entanglement. 
The renormalization of MPS~\cite{Sch11} and MERA~\cite{Vid07} can be seen as such schemes with low-depth circuits. \\ \hline
QTM-AQC & This is to use slow or external dynamical control to change MPS.
For MPS as ground states of a family of Hamiltonian $H(\lambda)$,
it can be used to make state conversions within a phase of matter or to study phase transition~\cite{CGW10}. \\ \hline
QTM-MBQC & Resource states using PEPS form are also widely studied in MBQC~\cite{GE07}.
Local projective measurements induce a local change of on-site tensors, 
which simulate a desired gate sequence on the ``virtual'' bond space.  \\ \hline
QCA-TVQC & Each primary step of a QCA can be seen as a TV operation, but a whole QCA is not.
In other words, QCA is a TVQC but with different transversality.
This relates to a code concatenation framework which is usually employed in QEC,
which combines TV gates with respect to each code being concatenated~\cite{NC00}. \\ \hline
QCA-MPQW & The original MPQW schemes have features of QCA, as has been discussed in Sec.~\ref{subsec:mpqw}. \\ \hline
QCA-TQC & There are QCA with topological features, such as non-trivial index~\cite{GNVW12,FH19,FHH19}.
For instance, 1D translation has a nontrivial index, although it does not change entanglement.
The index captures the imbalance of right and left-moving information. 
Topological QCA shall be fault tolerant against local noises,
while its application in quantum computing needs more study. \\ \hline
QCA-MBQC & An example is the recent study of using projective measurements to simulate global gates in QCA, 
such as Clifford QCA~\cite{SNB+19}. \\ \Xhline{3\arrayrulewidth}
    \end{tabular}
    \label{tab:combuqcm}
\end{table}

\subsection{Scheme II: Codes with FDLU gates}
\label{subsec:fdluc}

\begin{figure}[t!]
    \centering
    \includegraphics[width=.3\textwidth]{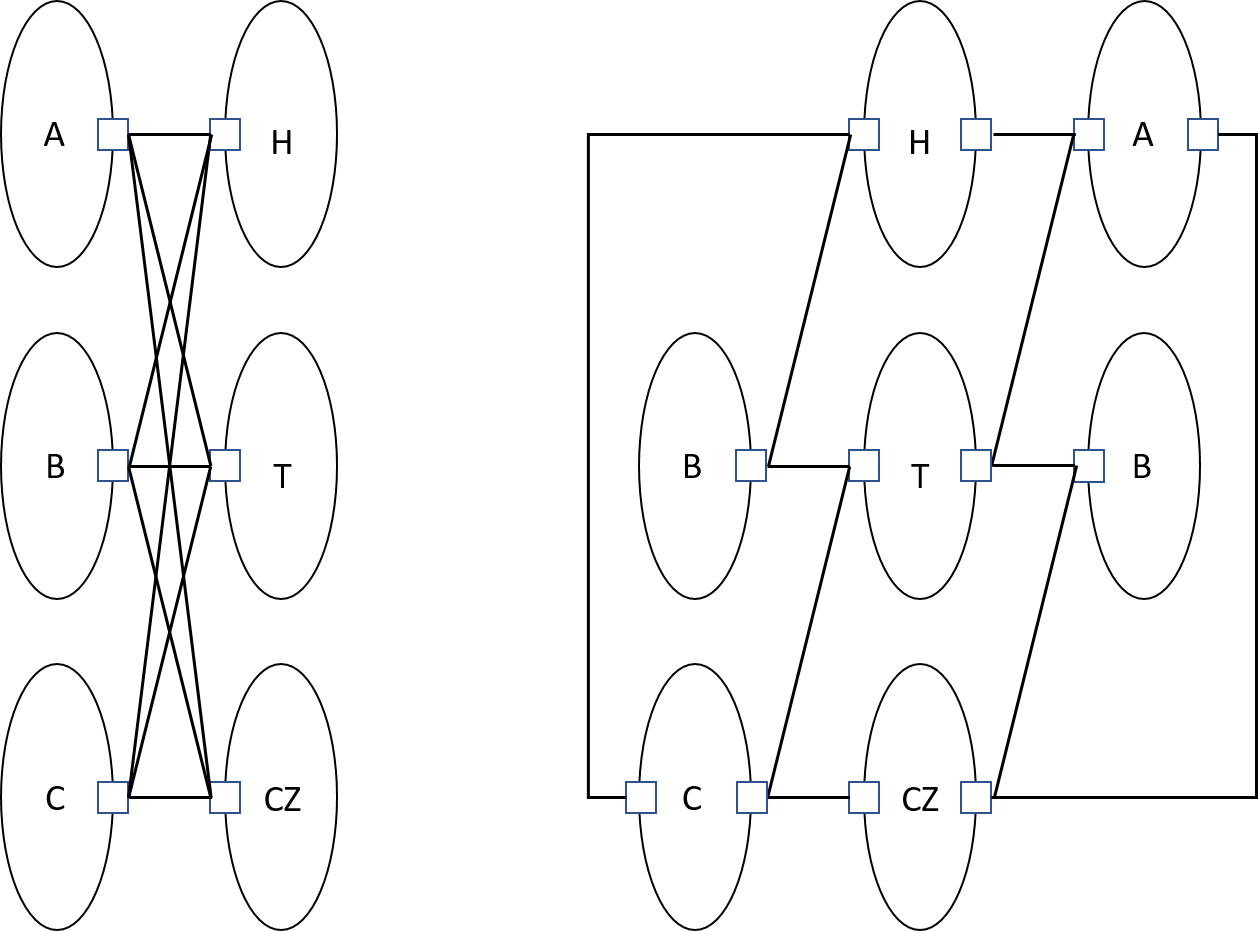}
    \caption{A variation of MPQW in QCM as an exemplary codes with FDLU gates. 
    Loops store gates H, T, CZ, or qubits (e.g., A, B, C), 
    the small boxes are switches, straight wires are for propagation.
    (Left): A scheme with the $K_{3,3}$ graph, which involves crossing wires.
    (Right): Crossing wires can be avoided by using a few loops to store a qubit (here, B),
    and this method extends to any finite number of qubits.}
    \label{fig:mpqwqcm}
\end{figure}

In QCM, gates are assisted with classical control. 
Here we describe a scheme, as a variation of MPQW, using interactions of excitations to realize FDLU gates. 
For MPQW, the initial wavepackets with special momentum are resources.
Interaction time is fixed by the graph length, wavepacket width, and its momentum.
This is actually an advantage of using discrete lattice to digitalize time.
A gate is stored in a local region of the lattice as a quantum oracle. 
Now, instead of using a global Hamiltonian $H$, 
we use three such gate regions for H, T, and CZ gates.
The H and T gate each can be stored in a 1D loop, 
while the CZ or CNOT gate can be stored in a pair of coupled loops.
The gates are executed by the dynamical Hamiltonian evolution.
To encode qubits,
we can use stationary local wavepackets or moving wavepackets stored in loops.
Assuming the ability to control momentum of wavepackets, 
they can be guided into gate regions and out,
with switches to open or close a loop. 
A sequence of gates are simulated by a sequence of controls, 
guiding wavepackets into and out of the gate regions.
A schematics is shown in Fig.~\ref{fig:mpqwqcm}.

This scheme is similar with those from Refs.~\cite{Nag10,Nag12,TGL+16} using 1D chains,
but our scheme is in QCM so classically controllable. 
There is no need to pick a spatial direction as the simulated time. 
From a similarity with photonic quantum computing,
even an ensemble of local wavepackets separated far enough on a single loop 
can be used for a qubit. 
There are challenges, though, such as the precise timing and change of momentum,
besides the challenge to realize such a system.
Two wavepackets need to be precisely guided into the coupled loops to interact 
to implement an entangling gate. 
To enhance the fault-tolerance, systems with topological or symmetry protection can be employed.
Despite of being dynamical instead of geometric or holonomic,
the logical gates are stored in the underlying lattices,
which can be protected by error correction using local Hamiltonian terms.


\subsection{Scheme III:  higher-order MPS}
\label{subsec:topQTM}

\begin{figure}[b!]
    \centering
    \includegraphics[width=.3\textwidth]{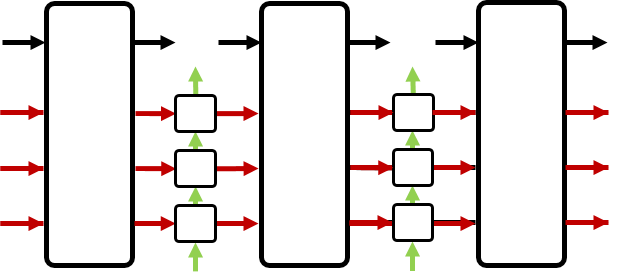}
    \caption{A quantum circuit to prepare a HO-MPS. 
    The colors are for different spaces: black for the physical space, 
    red for the 1st-order adversary space, 
    and green for the 2nd-order adversary space.
    The boxes are unitary gates.
    The arrows highlight the input and output.}
    \label{fig:HOMPS}
\end{figure}

When quantum information is expressed as a MPS, 
both the system and adversary can actually carry information. 
We can consider nontrivial operations on both of them. 
Here we study high-depth operations on the adversary of a MPS.

Now we introduce a higher-order MPS to capture the structure of the adversary.
Recall that a MPS contains a set of local matrices and boundary states.
When the bond dimension is large, 
we can further decompose the matrices $A$ as matrix-product operators (MPO), 
and boundary states as MPS.
For instance, if $\chi=d_1^{N_1}$, then the adversary space can be viewed as $N_1$ local systems of dimension $d_1$.
Decomposing the MPS matrices $A$ as  
\be A^{i}=\sum_{\vec{\mu},\vec{\nu}} \T tr(C^iB^i_{\mu_1\nu_1}\cdots B^i_{\mu_{N_1}\nu_{N_1}})|\vec{\mu}\ket \bra \vec{\nu}| \ee
lead to a second bond space with dimension $\chi_1$,
acted upon by the $B$ matrices and the boundary matrices $C^i$,
for $\vec{\mu}=(\mu_1,\dots,\mu_{N_1})$, $\vec{\nu}=(\nu_1,\dots,\nu_{N_1})$.
This can be carried over a step further if $\chi_1$ is large,
and this altogether defines higher-order MPS (HO-MPS).
Note it may appear similar with PEPS, 
but here tensors can share the same physical index.

Using the quantum circuit form of a MPS, 
this is equivalent to apply a MPU at each step, see Fig.~\ref{fig:HOMPS}.
The bond space of the MPU can be further decomposed if it is large. 
This can be viewed as a concatenation of operations on the adversary space. 
This formalism can be employed to 
describe high-depth such as topological operations on MPS which change entanglement significantly.

\subsection{Scheme IV: Controlled QCA}

We mentioned that the Hamiltonian for a QCA can be time-dependent.
This leads to a scheme of adiabatic QCA, or more generally, classically controlled QCA.
A concrete example is to use parallel local adiabatic controls to realize each primary operation in a QCA. 
The controls can be local in spacetime, namely, 
each control is on a local interaction term $h(s,n)$ centered around site $n$,
and the speed of the adiabatic passage $s$ is not a constant and is locally optimized. 
This makes it possible to use optimization algorithms to design a control program to solve given problems.
The output containing the solution can be encoded as bit strings of the qubit arrays,
which can be read out directly,
or as a probability distribution of the final state,
which requires multiple runs of the scheme.
Concrete examples and settings to manifest the power of a controlled QCA need further investigation. 

\section{Quantum algorithms}
\label{sec:qalg}

In this section, we study quantum algorithms in the light of computing models.
Many quantum algorithms are inspired by computing models,
although an algorithm can be realized in all equivalent UQCM.
An algorithm is designed to solve an instance of a problem.
Given a problem $P$, there is a presentation $E$ that translates the problem into objects in a model $M$, 
and the algorithm $A$ can be realized in the model $M$. 

For an algorithm $A$, there is actually another hidden algorithm $A_0$ that designs it. 
It is not hard to realize that $A_0$ also needs another algorithm to design it, and so on.
This sequence of algorithms continues in principle,
but we have to make a cut at some stage. 
The problem is given as input to $A_0$,
and the solution is the output of $A$. 
The nature of the two algorithms depends on the presentation $E$ of a problem.
Let the object be $U$. 
There are at least three types of presentations of it: 
\begin{itemize}
    \item bits $[U]$: this is used in a classical algorithm $A_0$ that takes $[U]$ as input,
        and outputs $[Q]$ as a classical description of a quantum algorithm $Q$.
     \item qubits $|U\ket$:  this is used in a quantum algorithm $Q_0$ that designs a quantum algorithm $Q$.
     This relates to SPQC.
    \item oracle $O_U$: this is used in a quantum algorithm $Q_0$ that contains another quantum algorithm $Q$. 
    This is a quantum meta-algorithm and relates to QSVT and quantum comb.
\end{itemize}

We shall note that there could also be other types of presentations that we do not study here. 
The first type is common and often used in QCM with classical description of problem.
However, it cannot be used when the presentation $[U]$ is not efficient. 
Below we study the other two types of algorithms with quantum oracles.
We propose a hybrid SPQC and draw connections between QSVT and higher-order quantum operations.



\subsection{General SPQC}
\label{subsec:spqc}

SPQC is the quantum analog of von Neumann architecture which stores programs in memory. 
Classical programs, as algorithms or functions, can be stored as bit strings, copied, or read out without disturbance. 
It turns out this is difficult for the quantum case. 
Usually, a program is a unitary quantum gate $U$. 
The problem of stored-program is to consider how to store $U$ in computing memory units. 
If there exists a unitary operator $G$ with 
\be G|\varphi\ket |P_U\ket= U|\varphi\ket |P_U'\ket,\ee
which means $G$ can extract the action of an unknown gate $U$ from a program state $|P_U\ket$ and apply on data state $|\varphi\ket$,
then $\bra P_U|P_V\ket=0$ for any different $U$ and $V$~\cite{NC97}. 
This means any tiny disturbance of $U$ leads to a state orthogonal with $|P_U\ket$, a kind of `orthogonal catastrophe'.
This is often referred as the no-programming theorem since this implies the program space dimension tends to infinity. 

Recently, a careful study~\cite{YRC20} shows that, with a given accuracy $\epsilon$ as input, the effective storage space for quantum programs scales efficiently with the accuracy,  although the system size of the program does not.
An optimal scheme is to build up the program state $|P_U\ket$ with $n$ identical usages of $U\in SU(d)$, 
with an inefficient scaling $n^2\sim \frac{1}{\epsilon}$,
which originates from the Heisenberg limit~\cite{TA14}. 
The scheme to extract $U$ is to perform a covariant global POVM on the $n$-qudit program state.


The SPQC applies to the setting that a client holds data state $|\varphi\ket$,
while an agent has the desirable program $U$ encoded as $|P_U\ket$, 
unknown to the client, and the goal is to prepare a final state $|\varphi_f\ket=U|\varphi\ket$.
The agent does not want to reveal the program $U$.
However, the setting can be modified to improve the efficiency.  
We find that the agent can choose a hybrid scheme of $|P_U\ket$,
and even reveal partial information to the client. 

We first analyze a totally classical scheme.
An efficient way to build up a classical storage of program is to store the information of the quantum circuit to simulate $U$, with a given accuracy $\epsilon$. 
The accuracy has to be given as input, since otherwise an arbitrary $U$ cannot be stored classically efficiently. With the given accuracy, one can first decompose $U$ as a sequence of gates from a universal gate set, e.g., $\{H, T, CZ\}$~\cite{DN06,Sel12,KMM13}. We can use two bits to store the gate types: 00 for $H$, $01$ for $T$, 10 for $CZ$. There is no need to store the gates themselves. 
We also need bits to store the spacetime location of a gate: 
where and when it acts. 
Given a random $U\in SU(d)$, the circuit size is of order $O(d^2 \log\frac{1}{\epsilon})$ acting on $\log d$ qubits. So a classical program can be prepared efficiently. 

The classical program does not have any quantum feature. 
So quantum computing with classical program needs the full ability of the client to perform quantum gates;
otherwise a client cannot use classical programs. 
Now we seek a hybrid approach.
In this regard, the magic-state injection model~\cite{BK05,Kni05} 
can be seen as examples. 
The program contains a classical part, which describes the gate sequence to simulate a given $U$ within accuracy $\epsilon$, 
and a quantum part made up of (copies of) magic states. 
The program has to be a white box, since otherwise it cannot be properly used. 

A general hybrid scheme would break a program into a few blocks and connections among them,
and store some blocks as quantum program states.
The program state $|P_U\ket$ contains classical bit strings as a part, 
hence orthogonal for different gates. 
This avoids the blow up of the program space dimension
since $k$ bits can store $2^k$ number of distinguishable bit strings.
The algorithm blocks can be of special functions, 
such as Clifford circuits, circuits with symmetry, quantum Fourier transform,
diagonal gates, or product of T gates. 
Depending on the contexts, easy gates or blocks can be stored classically,
while the hard ones require quantum programs. 
The client has partial quantum computing power such as performing Clifford circuits,
but the program states can still be blind.





\subsection{Quantum singular-value transformation}
\label{subsec:qsvt}

The QSVT is recently proposed as a unifying framework of some common quantum algorithms,
including Grover's search algorithm, quantum phase estimation, Hamiltonian evolution simulation, 
and some algorithms for matrix problems, etc~\cite{GSLW19,MRTC21}. 
Here we analyze QSVT from the point of view of quantum supermaps,
and discuss some extensions of QSVT.

QSVT is a generalization of quantum signal processing (QSP) on a qubit~\cite{LYC16}.
To realize any $U\in SU(2)$, QSP employs a fixed gate, $G$, 
and a set of phase gates of the form $Z(\phi)=e^{i\phi Z}$ to yield the approximation $\prod_k (Z(\phi_k)G)$.
The fixed gate $G$ can be given as an oracle, 
and the set of angles $\{\phi_k\}$ are the available resources.
Consider a space of matrices $M_d(\mathbb{C})$.
A matrix $A\in M_d(\mathbb{C})$ can be expressed as the set of singular values and singular vectors
\be A=W\Sigma V=\sum_i \sigma_i |w_i\ket \bra v_i|.\ee
Now a block-encoding of $A$ puts it on the upper left block of a unitary $U$ of dimension $2d$,
with $A:=\Tilde{\Pi}U \Pi$.
In the singular-vector basis, $U$ is a union of qubits each with a singular value $\sigma_i$
that can be manipulated by QSP. 
Then a generalized phase gate $e^{i \phi \Pi}$ or $e^{i \phi \Tilde{\Pi}}$ 
sandwiched in between $U$ and $U^\dagger$ forms the element for QSVT, 
and a set of angles $\{\phi_k\}$ can change $A$ to the desired form for the solution.
The use of $U$ and $U^\dagger$ is to convert between the left and right singular vectors.

As a matrix transformation, QSVT converts $A$ to $B=W\Sigma' V$ for
$A$ and $B$ with the same set of left and right singular vectors,
and $\Sigma'=\text{poly}(\Sigma)$ as a certain polynomial function of $\Sigma$.
Denote such $B$ as $f_\text{sv}(A)$.
Now, given any $A$ and $B$, is there a QSVT between them?
We did not find such a solution in general, 
but there is a nontrivial extension of QSVT using linear combination of unitary (LCU) algorithms~\cite{Long06,Long11,CW12}.
Given a set $\{A_i\}$, the conversion $\{A_i\} \mapsto B$ for 
\be B=\sum_i \beta_i f_\text{sv}(A_i) \label{eq:sumf}\ee
can be realized by LCU with QSVT for each $A_i$. 
The `seed' operators $A_i$ do not need to commute with each other, $[A_i,A_j]\neq 0$.
The Hamiltonian evolution simulation $e^{iHt}$ is a special case of this:
the cosine term $\cos Ht$ and sine term $\sin Ht$ commute.



QSVT can be seen as an example of quantum comb, which acts on the data system and an adversary. 
In other words, the quantum comb can be treated as an extension of QSVT, 
which takes quantum operations (or even quantum combs) as input (oracle),
and, in particular, it has been used for more general tasks such as quantum channel discrimination~\cite{CAP08}.
For QSVT, the input is $U$ and $U^\dagger$, and
there is no entanglement between the data and adversary. 
The adversary performs phase kickback via QSP. 
The linear combination of QSVTs can also be treated in the framework of quantum comb
by enlarging the dimension of the adversary.
Therefore, motivated by the study above, it is highly desirable to seek new quantum algorithms along this line. 




\section{Conclusion}
\label{sec:conc}

In this work, we attempted to understand a few universal quantum computing models
from a unifying physical framework. 
The study turns out to be fruitful.
On the one hand, we arrange them onto a table as two categories;
on the other hand, we find promising schemes that have not been studied so far.
We hope our work can benefit the study of modelings, codes, or algorithms. 
Below we discuss a few issues that are largely omitted in this work.

There are non-universal models
which often can be viewed as restrictions of universal ones.
The category-I models mainly refer to universality,
and the category-II models mainly refer to fault tolerance,
but universality also implicitly requires the latter since the logical identity gate is needed.
We could replace the two columns of the table by non-universal ones,
and replace error-correction by other decoherence-engineering techniques.
This is useful for the algorithmic quantum-classical separation,
for classification of special circuits, states, or dynamics, 
or for the demonstration of quantum advantages in the NISQ era~\cite{Pre18}.



Our study stays on the operator level, namely,
we do not consider measurement readout outcomes such as probability distribution. 
The universality defined in this work mainly refers to quantum state conversion,
but the readout from the final state could be nontrivial.
Taking this into account leads to other algorithms or tasks,
such as sampling~\cite{AA11}, quantum-classical hybrid algorithms~\cite{FGG14}, 
etc that are capable to show quantum advantages.

Finally, on the ``super-operator'' level
we studied the structure of quantum meta-algorithms using the framework of quantum comb,
which also relates to quantum Turing machines.
We expect this can be used to study fully-quantum machine learning
and quantum optimization algorithms,
which often involve the tuning of meta-parameters or meta-operators~\cite{DB18,VPB18,MBW+19,BLSF19}. 
The quantum adversary is deployed as a resource to speedup the computation on the system, 
whose quantum advantages have not been fully explored.






\section*{Acknowledgements}

The author thanks G. Long, T. Lan, R. Raussendorf, and Y. Yang for discussions and suggestions. 
This work has been supported by the National Natural Science Foundation of China 
(Grant No. 12047503).

{\footnotesize
\bibliography{ext}{}
\bibliographystyle{ieeetr}
}
\end{document}